\documentclass[12pt]{article}
\usepackage{epsf}
\usepackage[dvips]{color}
\usepackage{exscale}
\usepackage[dvips]{graphicx}
\usepackage{amsmath}
\usepackage{amssymb}
\usepackage{latexsym}
\usepackage{url}

\begin{document}
\setlength{\baselineskip}{18pt}
\begin{titlepage}
\begin{flushright}
\begin{tabular}{l}
EPHOU-13-004
\end{tabular} 
\end{flushright}

\vspace*{1.2cm}
\begin{center}
{\Large\bf Constraints on neutrino mass ordering and degeneracy from Planck and neutrino-less double beta decay}
\end{center}
\lineskip .75em
\vskip 1.5cm

\begin{center}
{\large Naoyuki Haba$^{1,2}$ and Ryo Takahashi$^2$}\\

\vspace{1cm}

$^1${\it Graduate School of Science and Engineering, Shimane University, 

Matsue 690-8504, Japan}\\

$^2${\it Department of Physics, Faculty of Science, Hokkaido University, 

Sapporo 060-0810, Japan}\\

\vspace*{10mm}
{\bf Abstract}\\[5mm]
{\parbox{13cm}{\hspace{5mm}
We investigate constraints on neutrino mass ordering and degeneracy by considering the first 
cosmological result based on {\it Planck} measurements of the cosmic microwave 
background. It is shown that the result at 95\% CL rejects a neutrino mass degeneracy larger than 85\% (82.5\%) for the normal (inverted) hierarchical case. We can also find some regions where the neutrino mass ordering will be able to be distinguished by combining a value of sum of the neutrino masses with an effective neutrino mass determined by neutrino-less double beta decay experiments. The results are obtained from the latest data of neutrino oscillation, cosmic microwave background, and the neutrino-less double beta decay experiments. These have significance in the discrimination of the neutrino mass ordering.\\

}}
\end{center}
\end{titlepage}

\section{Introduction}

Neutrino oscillation experiments have established that neutrinos have tiny masses compared to other standard model (SM) fermion masses. Further, recent precision measurements of mixing angles in the Pontecorvo-Maki-Nakagawa-Sakata (PMNS) matrix~\cite{Maki:1962mu} have clarified that there are two large ($\theta_{12}$ and $\theta_{23}$) and one small ($\theta_{13}$) mixing angles~\cite{An:2012eh,tortola,GonzalezGarcia:2012sz}. Regarding the neutrino masses, only two mass squared differences are determined. Therefore, one can consider a normal mass hierarchy (NH: $m_1<m_2<m_3$) or an inverted one (IH: $m_3<m_1<m_2$) where $m_i$ are mass eigenvalues of the three light neutrinos. A determination of the neutrino mass ordering is one of important tasks in neutrino physics.

Recently, an important result has just been reported by {\it Planck} measurements of the cosmic microwave background (CMB)~\cite{Ade:2013zuv}, which is
 \begin{eqnarray}
  \sum m_\nu<0.230\mbox{ eV ({\it Planck}$+$WP$+$highL$+$BAO)},
 \end{eqnarray}
for the sum of the three light neutrino masses with an 
assumption of three species of degenerate neutrinos. An upper limit for the sum
 of the neutrino masses is important to determine the neutrino mass ordering and
 constrain a degeneracy among the neutrino masses. In this letter, we 
investigate the first cosmological result based on the {\it Planck} CMB 
measurement for the neutrino mass ordering and the neutrino mass degeneracy.

Regarding the neutrino mass ordering, values of an effective neutrino mass for the neutrino-less double beta decay  ($0\nu\beta\beta$) are differently predicted in the NH and IH cases. Therefore, it is also important to distinguish the neutrino mass ordering if the neutrinos are Majorana particles. We will consider the latest result for an effective neutrino mass from a $0\nu\beta\beta$ experiment and an expected reach for the mass by future $0\nu\beta\beta$ experiments with the cosmological constraint on the sum of the neutrino masses for the discrimination of the neutrino mass ordering. Our results will be obtained from the latest data of neutrino oscillation, CMB, and $0\nu\beta\beta$ experiments.

\section{Constraints on neutrino mass ordering from Planck and neutrino-less 
double beta decay}

The neutrino oscillation experiments determine only two mass squared differences
 of neutrinos, $\Delta m_{21}^2$ and $\Delta m_{31}^2$ (or $\Delta m_{32}^2$) 
defined as
 \begin{eqnarray}
  \Delta m_{21}^2 &\equiv& m_2^2-m_1^2, \\
  \Delta m_{31}^2 &\equiv& m_3^2-m_1^2~~~\mbox{ for the NH}, \\
  \Delta m_{32}^2 &\equiv& m_3^2-m_2^2~~~\mbox{ for the IH}.
 \end{eqnarray}
Therefore, the neutrino mass spectrum can be described by the two mass squared differences and 
a remaining mass as
 \begin{eqnarray}
  && (m_1,m_2,m_3) \\
  && =(m_1,\sqrt{\Delta m_{21}^2+m_1^2},\sqrt{\Delta m_{31}^2+m_1^2})\mbox{ or }
      (\sqrt{m_3^2-\Delta m_{31}^2},\sqrt{\Delta m_{21}^2-\Delta m_{31}^2+m_3^2}
       ,m_3), \nonumber 
 \end{eqnarray}
for the NH, and
 \begin{eqnarray}
  && (m_1,m_2,m_3)=(m_1,\sqrt{\Delta m_{21}^2+m_1^2},
       \sqrt{\Delta m_{32}^2+\Delta m_{21}^2+m_1^2}) \nonumber \\ 
  && \phantom{(m_1,m_2,m_3)=}\mbox{ or }
      (\sqrt{m_3^2-\Delta m_{32}^2-\Delta m_{21}^2},\sqrt{m_3^2-\Delta m_{32}^2}
       ,m_3), 
 \end{eqnarray}
for the IH. The values of the neutrino mass squared differences are determined as
 \begin{eqnarray}
  \Delta m_{21}^2 &=& 7.50_{-0.50}^{+0.59}\times10^{-5}\mbox{ eV}^2, \label{21} \\
  \Delta m_{31}^2 &=& 2.47_{-0.20}^{+0.22}\times10^{-3}\mbox{ eV}^2, \\
  \Delta m_{32}^2 &=& -2.43_{-0.22}^{+0.19}\times10^{-3}\mbox{ eV}^2, \label{32}
 \end{eqnarray}
at $3\sigma$ level~\cite{GonzalezGarcia:2012sz}.
 
\begin{figure}
\hspace{4.4cm}(a)\hspace{7.6cm}(b)
\begin{center}
\includegraphics[scale=0.89]{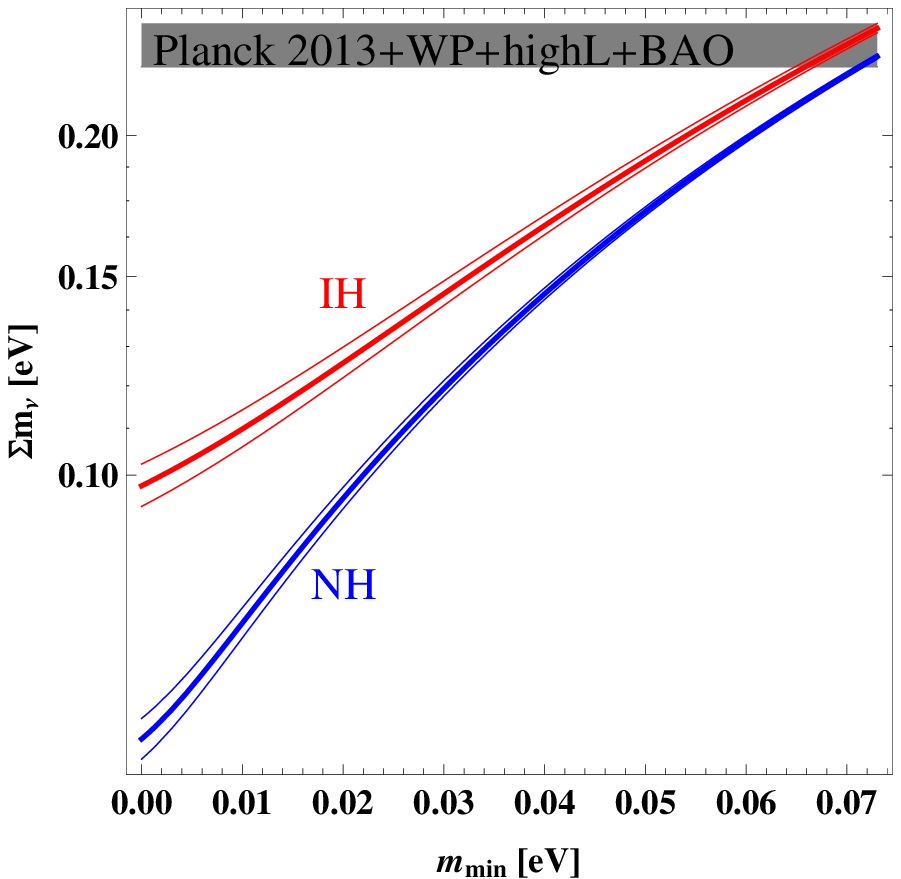}
\includegraphics[scale=0.89,bb=0 0 260 260]{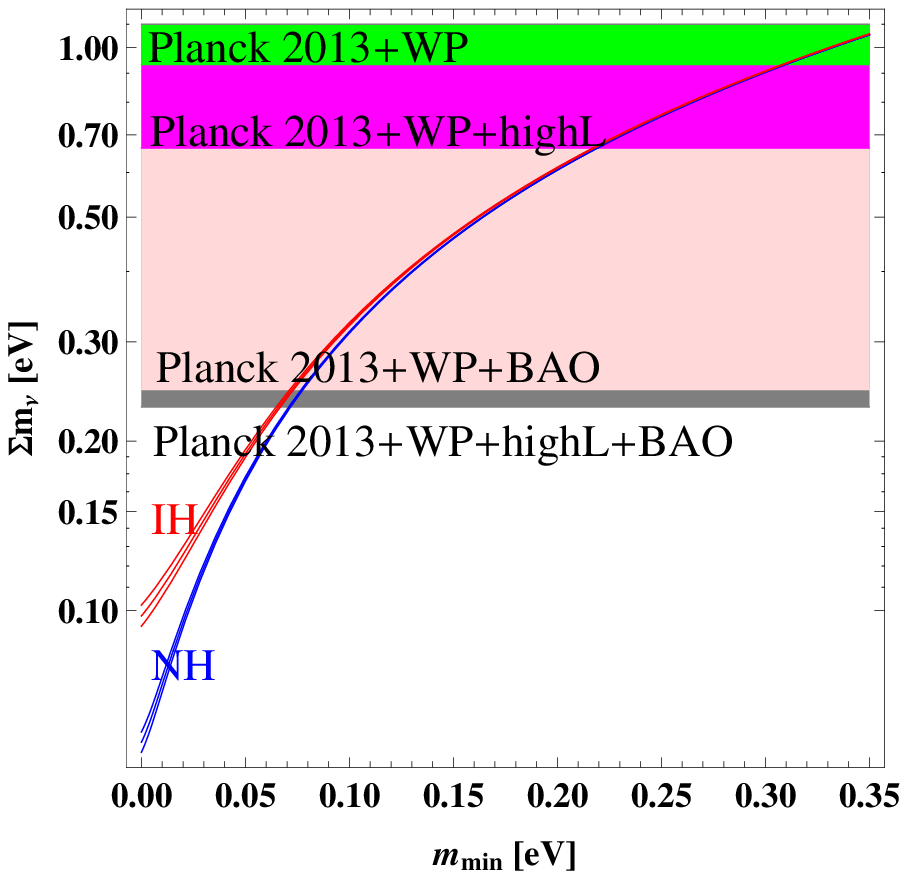}
\end{center}
\hspace{4.4cm}(c)\hspace{7.6cm}(d)
\begin{center}
\includegraphics[scale=0.89]{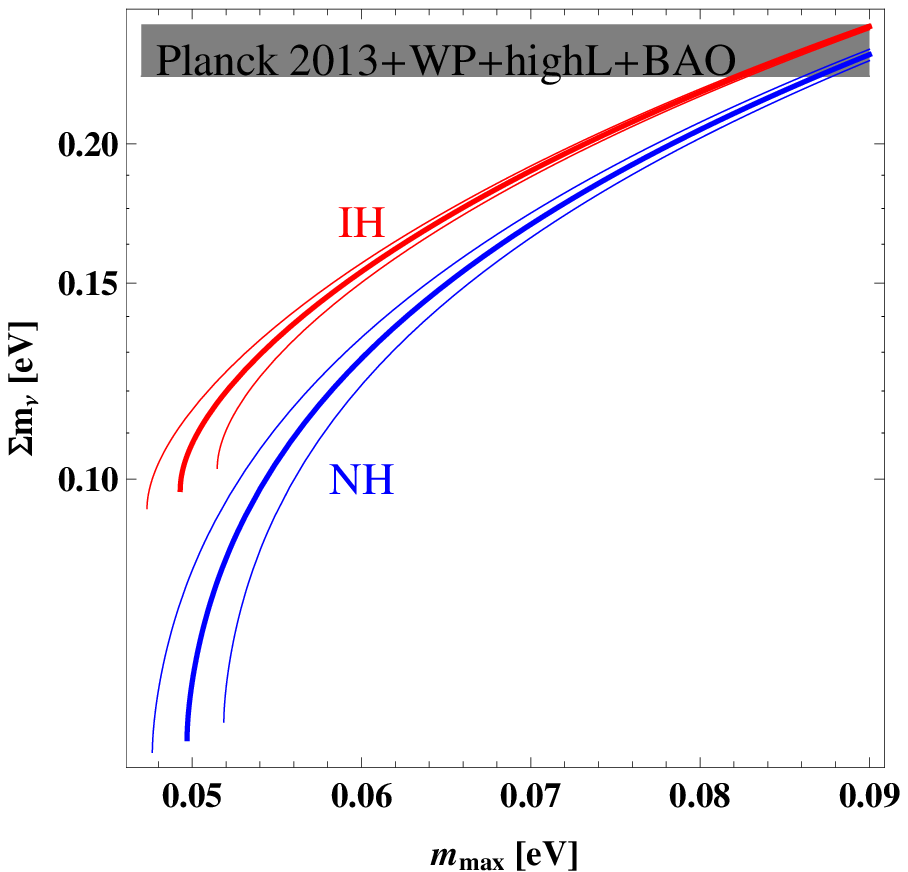}
\includegraphics[scale=0.89]{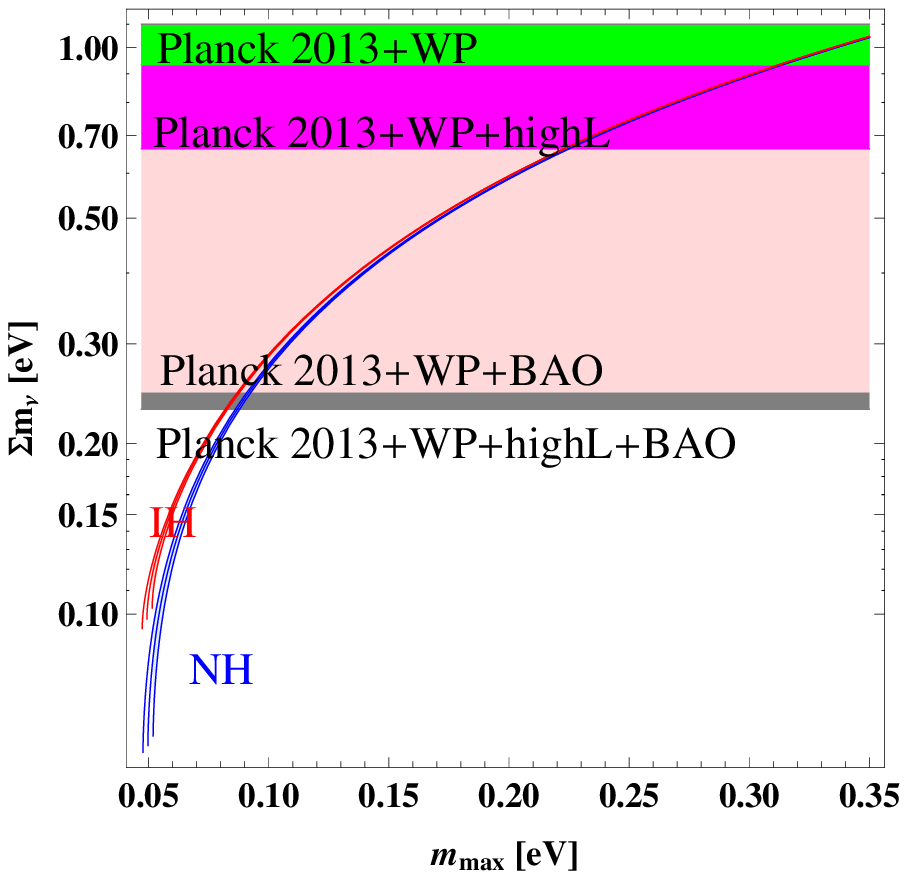}
\end{center}
\caption{The sum of the neutrino masses in functions of $m_{\rm min}$ and $m_{\rm max}$, and cosmological bounds on the sum of the neutrino mass from the {\it Planck} with other data.}
\label{fig1}
\end{figure}

\begin{figure}
\hspace{4.4cm}(a)\hspace{7.6cm}(b)
\begin{center}
\includegraphics[scale=0.89]{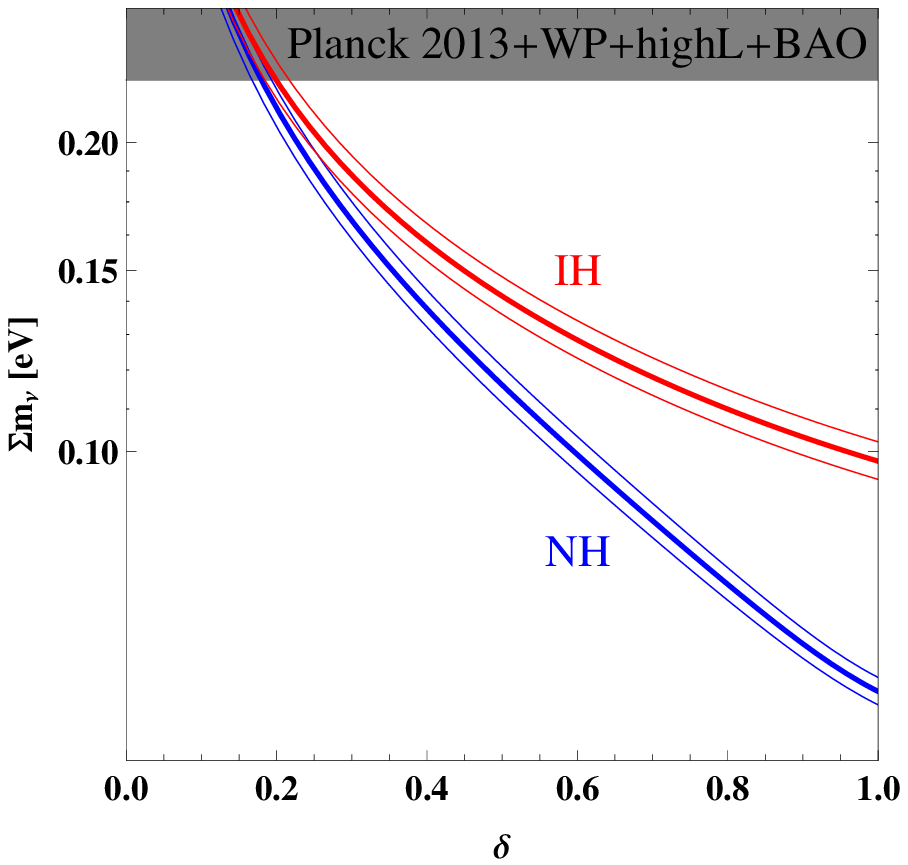}
\includegraphics[scale=0.89]{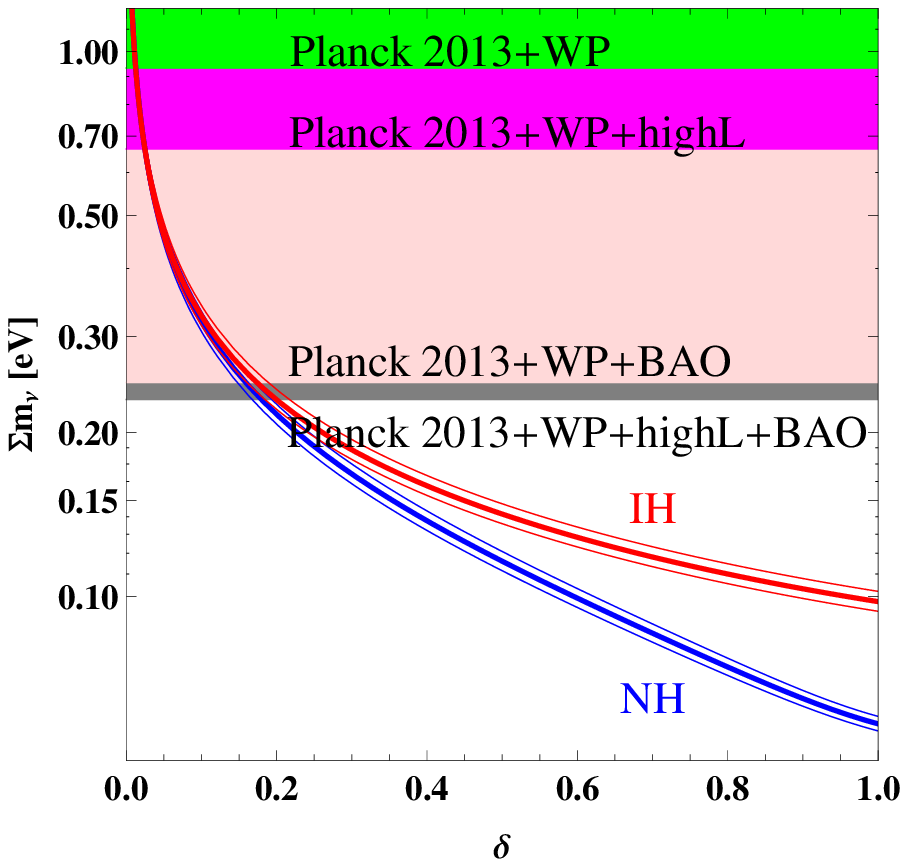}
\end{center}
\caption{The sum of the neutrino masses in function of the neutrino mass degeneracy $\delta$, and cosmological bounds on the sum of the neutrino mass from the {\it Planck} with other data.}
\label{fig2}
\end{figure}

On the other hand, the first cosmological result based on {\it Planck} 
measurements of the CMB~\cite{Ade:2013zuv} have presented an upper bound on the sum of the neutrino masses assuming no extra relics with a WMAP polarization low-multipole 
likelihood at $\ell\leq23$ (WP)~\cite{Bennett:2012fp,Planck:2013kta}, 
high-resolution (highL) CMB data, and baryon acoustic oscillation (BAO) surveys as,
 \begin{eqnarray}
  \sum m_\nu<0.230\mbox{ eV ({\it Planck}$+$WP$+$highL$+$BAO)}, \label{lim}
 \end{eqnarray}
at 95\% CL. We show this upper limit in Fig.~\ref{fig1} and \ref{fig2}, in which
 the horizontal axes are $m_{\rm min}$ or $m_{\rm max}$ in Fig.~\ref{fig1} 
(see also \cite{Lesgourgues:2006nd} for a plot in 
($m_{\rm min}$,$\sum m_\nu$) plane with old data), and $\delta$ in 
Fig.~\ref{fig2}.
$m_{\rm min}$ and $m_{\rm max}$ stand for minimal and maximal values among three neutrino mass eigenvalues, respectively. Therefore, these are defined as
 \begin{eqnarray}
  &&m_{\rm min}\equiv m_1,\qquad m_{\rm max}\equiv m_3\qquad\mbox{for the NH}, \\
  &&m_{\rm min}\equiv m_3,\qquad m_{\rm max}\equiv m_2\qquad\mbox{for the IH},
 \end{eqnarray}
respectively. With these definitions, the sum of the neutrino masses are described by
 \begin{eqnarray}
  \sum m_\nu=
   \left\{
    \begin{array}{l}
     m_{\rm min}+\sqrt{\Delta m_{21}^2+m_{\rm min}^2}+\sqrt{\Delta m_{31}^2+m_{\rm min}^2} \\
     \sqrt{m_{\rm max}^2-\Delta m_{31}^2}+\sqrt{m_{\rm max}^2-\Delta m_{31}^2+\Delta m_{21}^2}+m_{\rm max}
    \end{array}
   \right.,
 \end{eqnarray}
for the NH, and
 \begin{eqnarray}
  \sum m_\nu=
   \left\{
    \begin{array}{l}
     \sqrt{m_{\rm min}^2-\Delta m_{32}^2-\Delta m_{21}^2}+\sqrt{m_{\rm min}^2-\Delta m_{32}^2}+m_{\rm min} \\
     \sqrt{m_{\rm max}^2-\Delta m_{21}^2}+m_{\rm max}+\sqrt{m_{\rm max}^2+\Delta m_{32}^2} 
    \end{array}
   \right.,
 \end{eqnarray}
for the IH in Fig~\ref{fig1}. $\delta$ indicates a magnitude degeneracy of the neutrino masses defined by
 \begin{eqnarray}
  \delta\equiv\frac{m_{\rm max}-m_{\rm min}}{m_{\rm max}}. \label{de}
 \end{eqnarray}
Therefore, the limit of $m_{\rm min}\rightarrow0~(m_{\rm max})$ means $0\%~(100\%)$ of degeneracy among the neutrino masses, respectively. In Fig.~\ref{fig1} and \ref{fig2}, thick curves for both the NH and the IH are the sum of the neutrino masses with the best fit values for two mass squared differences. The lower and upper curves from the thick curve are also given by using values of two mass squared differences in $3\sigma$ range \eqref{21}-\eqref{32}. In Fig.~\ref{fig1} (a), (c) and Fig.~\ref{fig2} (a), the upper limit \eqref{lim} is shown by (gray) shaded region. In Fig.~\ref{fig1} (b), (d) and Fig.~\ref{fig2} (d), other limits at 95\% CL as
 \begin{eqnarray}
  \sum m_\nu<
   \left\{
    \begin{array}{ll}
     0.247\mbox{ eV} &  \mbox{ ({\it Planck}$+$WP$+$BAO)} \\
     0.663\mbox{ eV} &  \mbox{ ({\it Planck}$+$WP$+$highL)} \\
     0.933\mbox{ eV} &  \mbox{ ({\it Planck}$+$WP)}
    \end{array}
   \right.,
 \end{eqnarray}
are also shown by (light red, magenta, and green) shaded regions, respectively. One can easily find from all figures that if cosmologically observed value of the sum of the neutrino masses is smaller than a minimal value of the sum in the IH, $\sum m_\nu<\mbox{min}[\sum m_\nu|_{m_3=0}]\simeq0.0987$ eV within $3\sigma$ range, the IH of the neutrino mass spectrum can be ruled out.

One can also note that if the sum of the neutrino masses could be determined within a region of $\mbox{min}[\sum m_\nu|_{m_3=0}]\simeq0.0987\mbox{ eV}\leq\sum m_\nu<0.23$ eV in future, a value of $m_{\rm min}$ or $m_{\rm max}$ can determine the neutrino mass ordering. For instance, $\sum m_\nu=0.2$ eV limits $m_{\rm min}$ and $m_{\rm max}$ to $0.060$ eV$\lesssim m_{\rm min}\lesssim0.062$ eV, $0.078$ eV$\lesssim m_{\rm max}\lesssim0.080$ eV for the NH and $0.052$ eV$\lesssim m_{\rm min}\lesssim0.056$ eV, $0.072$ eV$\lesssim m_{\rm max}\lesssim0.074$ eV for the IH, respectively. An experiment using atoms or molecules with an atomic process of radiative emission of neutrino pair (RENP) for neutrino spectroscopy might give a constraint on the absolute neutrino mass and/or mass ordering, or might independently determine them~\cite{Fukumi:2012rn}.

{}Fig.~\ref{fig2} shows the sum of the neutrino masses in function of the neutrino mass degeneracy $\delta$, and cosmological bounds on the sum of the neutrino mass from the {\it Planck} with other data. The meanings of curves and shaded regions are the same as in Fig.~\ref{fig1}. One can replace $m_{\rm min}$ or $m_{\rm max}$ in the sum of the neutrino masses by $\delta$ defined in \eqref{de}. We find that a magnitude of degeneracy $\delta\lesssim0.15~(0.175)$ is ruled out at 95\% CL for the NH (IH). This means that a degeneracy larger than about 85 (82.5)\% is rejected for the NH (IH).

\begin{figure}
\begin{center}
\includegraphics[scale=1]{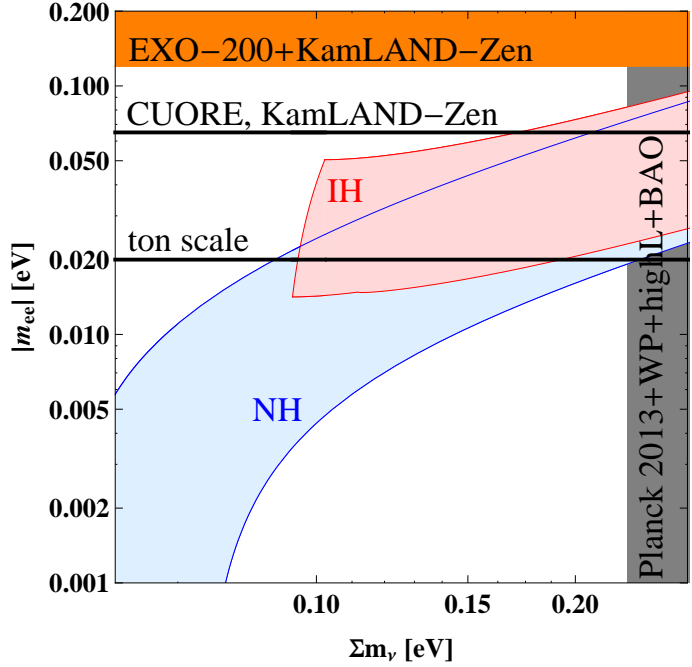}
\end{center}
\caption{The cosmological constraint on the sum of the neutrino mass from the {\it Planck} with a result from the neutrino-less double beta decay ($0\nu\beta\beta$) experiment.}
\label{fig3}
\end{figure}

Finally, in Fig.~\ref{fig3}, we compare the cosmological constraint on the sum of the neutrino mass from the {\it Planck} with a result from the $0\nu\beta\beta$ experiment, which constrain an effective neutrino mass defined as~(e.g., see~\cite{Vissani:1999tu}),
 \begin{eqnarray}
  |m_{ee}|\equiv\left|\sum_iU_{ei}^2m_i\right|=\left|c_{12}^2c_{13}^2m_1+s_{12}^2c_{13}^2m_2e^{2i\alpha}+s_{13}^2m_3e^{2i\beta}\right|, \label{mee}
 \end{eqnarray}
with $s_{ij}\equiv\sin\theta_{ij}$ and $c_{ij}\equiv\cos\theta_{ij}$ where $U$ is the PMNS matrix, $\theta_{ij}$ are the mixing angles in the PMNS matrix, $\alpha$ is one of Majorana phases, and $\beta$ is a re-defined CP phase by the Dirac CP phase ($\delta_D$) and another Majorana one ($\beta'$) as $\beta\equiv\beta'-\delta_D$ (see also \cite{Fogli:2004as} for a plot in $(\sum m_\nu,|m_{ee}|)$ plane with old data).
The combined result from the EXO-200 and KamLAND-Zen experiments is 
$|m_{ee}|<(120-250)$ meV at 90\% CL~\cite{Gando:2012zm}.\footnote{See 
also~\cite{Faessler:2013hz} for the correlated uncertainties associated to the 
nuclear matrix elements of $0\nu\beta\beta$ within the quasiparticle random 
phase approximation.} It is shown by an orange shaded region in Fig.~\ref{fig3}.
 We also present an expected value of 65 meV (sensitivity of the 
CUORE~\cite{Arnaboldi:2002du} and KamLAND-Zen experiments at 90\% CL after a 5 
years exposures~\cite{GomezCadenas:2011it}) and 20 meV (ton scale experiment), 
which are described by thick horizontal lines in addition to the {\it Planck} 
limit in the vertical direction (the sum of the neutrino masses). It is known 
that minimal and maximal values of $|m_{ee}|$ for both the NH and IH cases are 
determined by differences of relative sings among each term in \eqref{mee}, 
which depend on Majorana phases (see e.g.~\cite{Lindner:2005kr}). The upper 
(maximal) and lower (minimal) boundaries of the region for both the NH and IH in
 particular correspond to
 \begin{eqnarray}
  |m_{ee}|=
   \left\{
    \begin{array}{ll}
     |c_{12}^2c_{13}^2m_1+s_{12}^2c_{13}^2m_2+s_{13}^2m_3| & \mbox{for the upper boundary in the NH and IH} \\
     |c_{12}^2c_{13}^2m_1-s_{12}^2c_{13}^2m_2-s_{13}^2m_3| & \mbox{for the lower boundary in the NH and IH}
    \end{array}
   \right..
 \end{eqnarray}
We also take mixing angles and two mass squared differences as
 \begin{eqnarray}
  (s_{12}^2,s_{13}^2,\Delta m_{21}^2,\Delta m_{31}^2)=
   \left\{
   \begin{array}{l}
    (0.34,0.030,7.00\times10^{-5}\mbox{ eV},2.27\times10^{-3}\mbox{ eV}) \\
    (0.34,0.030,7.00\times10^{-5}\mbox{ eV},2.69\times10^{-3}\mbox{ eV})
   \end{array},
   \right.
 \end{eqnarray}
for the upper and lower boundaries in the NH, respectively, and
 \begin{eqnarray}
  (s_{12}^2,s_{13}^2,\Delta m_{21}^2,\Delta m_{31}^2)=
  \left\{
   \begin{array}{l}
    (0.34,0.016,7.00\times10^{-5}\mbox{ eV},-2.65\times10^{-3}\mbox{ eV}) \\
    (0.34,0.030,7.00\times10^{-5}\mbox{ eV},-2.65\times10^{-3}\mbox{ eV})
   \end{array},
   \right.
 \end{eqnarray}
for the upper and lower boundaries in the IH, respectively. These values are 
marginal ones at $3\sigma$ level~\cite{GonzalezGarcia:2012sz}. The relative 
signs are obtained by taking the corresponding CP phase as $0$ or $\pi/2$, 
respectively. It can be seen from Fig.~\ref{fig3} that expected value by the 
CUORE and KamLAND-Zen experiment $|m_{ee}|=65$ meV 
cannot rule out the IH. 
However, if one combines a result for a value of $|m_{ee}|$ with one of 
$\sum m_\nu$, there are some regions in which one can distinguish between the NH
 and the IH. For instance, on the line of $|m_{ee}|=20$ meV, which may be reached by a ton scale experiment as discussed in~\cite{TheNEXT:2013lta}, with $0.19$ eV$\lesssim\sum m_\nu\lesssim0.23$ eV (or $\sum m_\nu\lesssim0.0987$ eV), the IH
 can be rejected. In a region of $0.023$ eV$\lesssim|m_{ee}|\lesssim0.080$ eV 
and $0.0987\mbox{ eV}\lesssim\sum m_\nu<0.23$ eV, there exists a region in 
which the only IH can be allowed. Since both the $0\nu\beta\beta$ experiments 
and cosmological CMB observation will come to an interesting region, a combining
 analysis will also become important to distinguish the neutrino mass ordering.

\section{Summary}

We studied constraints on the neutrino mass ordering and neutrino mass degeneracy by considering the first cosmological result based on the {\it Planck} measurements of the CMB. First, we shown the sum of the neutrino masses in functions of $m_{\rm min}$ and $m_{\rm max}$, and cosmological bounds on the sum of the neutrino mass from the {\it Planck} with other data. It was found that if cosmologically observed value of the sum of the neutrino masses is smaller than a minimal value of the sum for the IH case, $\sum m_\nu<\mbox{min}[\sum m_\nu|_{m_3=0}]\simeq0.0987$ eV within $3\sigma$ range, the IH of the neutrino mass spectrum can be ruled out. We could also found that if the sum of the neutrino masses could be determined within a region of $0.0987\mbox{ eV}\leq\sum m_\nu<0.23$ eV in future, a determination of value of $m_{\rm min}$ or $m_{\rm max}$ can clarify the neutrino mass spectrum. For instance, $\sum m_\nu=0.2$ eV limits $m_{\rm min}$ and $m_{\rm max}$ to $0.060$ eV$\lesssim m_{\rm min}\lesssim0.062$ eV, $0.078$ eV$\lesssim m_{\rm max}\lesssim0.080$ eV for the NH and $0.052$ eV$\lesssim m_{\rm min}\lesssim0.056$ eV, $0.072$ eV$\lesssim m_{\rm max}\lesssim0.074$ eV for the IH, respectively.

Next, we showed the sum of the neutrino masses in function of the neutrino mass degeneracy $\delta$, and cosmological bounds on the sum of the neutrino mass from the {\it Planck} with other data. It was found that a magnitude of degeneracy $\delta\lesssim0.15~(0.175)$ is ruled out at 95\% CL for the NH (IH). This means that a degeneracy larger than about 85 (82.5)\% is rejected for the NH (IH).

Finally, we compared the cosmological constraint on the sum of the neutrino mass
 from the {\it Planck} with a result from the $0\nu\beta\beta$ experiment, which
 constrains an effective neutrino mass. It was found that if one combines a 
result for a value of $|m_{ee}|$ with one of $\sum m_\nu$, there are some 
regions in which one can distinguish the NH and IH cases. For instance, on the line of $|m_{ee}|=20$ meV with $0.19$ eV$\lesssim\sum m_\nu\lesssim0.23$ eV (or $\sum m_\nu\lesssim0.0987$ eV), the IH can be rejected. In a region of $0.023$ eV$\lesssim|m_{ee}|\lesssim0.080$ eV and $0.0987\mbox{ eV}\lesssim\sum m_\nu<0.023$ eV, there exists a region in which the only IH can be allowed. 

Our results were obtained from the latest data of the neutrino oscillation, CMB, and $0\nu\beta\beta$ experiments. Since both the $0\nu\beta\beta$ experiments and cosmological CMB observation will come to an interesting region, a combining analysis will also become important to distinguish the neutrino mass ordering.

\subsection*{Acknowledgement}
This work is partially supported by Scientific Grant by Ministry of Education 
and Science, Nos. 00293803, 20244028, 21244036, 23340070, and by the SUHARA Memorial Foundation. The work of R.T. is supported by Research Fellowships of the Japan Society for the Promotion of Science for Young Scientists.



\begin{thebibliography}{99}
\bibitem{Maki:1962mu}
  Z.~Maki, M.~Nakagawa and S.~Sakata,
  Prog.\ Theor.\ Phys.\  {\bf 28} (1962) 870;
  B.~Pontecorvo,
  Sov.\ Phys.\ JETP {\bf 26} (1968) 984
  [Zh.\ Eksp.\ Teor.\ Fiz.\  {\bf 53} (1967) 1717].
  
\bibitem{An:2012eh}
  K.~Abe {\it et al.}  [T2K Collaboration],
  Phys.\ Rev.\ Lett.\  {\bf 107} (2011) 041801,
  [arXiv:1106.2822 [hep-ex]];
  P.~Adamson {\it et al.}  [MINOS Collaboration],
  Phys.\ Rev.\ Lett.\  {\bf 107} (2011) 181802
  [arXiv:1108.0015 [hep-ex]];
  Y.~Abe {\it et al.}  [DOUBLE-CHOOZ Collaboration],
  Phys.\ Rev.\ Lett.\  {\bf 108} (2012) 131801
  [arXiv:1112.6353 [hep-ex]];
  F.~P.~An {\it et al.}  [DAYA-BAY Collaboration],
  arXiv:1203.1669 [hep-ex];
  J.~K.~Ahn {\it et al.}  [RENO collaboration],
  arXiv:1204.0626 [hep-ex].

\bibitem{tortola}
  M.~Tortola, J.~W.~F.~Valle, and D.~Vanegas, arXiv:1205.4018 [hep-ph];
  G.~L.~Fogli, E.~Lisi, A.~Marrone, D.~Montanino, A.~Palazzo and A.~M.~Rotunno,
  Phys.\ Rev.\ D {\bf 86} (2012) 013012
  [arXiv:1205.5254 [hep-ph]].

\bibitem{GonzalezGarcia:2012sz}
  M.~C.~Gonzalez-Garcia, M.~Maltoni, J.~Salvado and T.~Schwetz,
  JHEP {\bf 1212} (2012) 123
  [arXiv:1209.3023 [hep-ph]].
    
\bibitem{Ade:2013zuv}
  P.~A.~R.~Ade {\it et al.}  [Planck Collaboration],
  arXiv:1303.5076 [astro-ph.CO].

\bibitem{Bennett:2012fp}
  C.~L.~Bennett, D.~Larson, J.~L.~Weiland, N.~Jarosik, G.~Hinshaw, N.~Odegard, K.~M.~Smith and R.~S.~Hill {\it et al.},
  arXiv:1212.5225 [astro-ph.CO].

\bibitem{Planck:2013kta}
  P.~A.~R.~Ade {\it et al.}  [Planck Collaboration],
  arXiv:1303.5075 [astro-ph.CO].

\bibitem{Lesgourgues:2006nd}
  J.~Lesgourgues and S.~Pastor,
  Phys.\ Rept.\  {\bf 429} (2006) 307
  [astro-ph/0603494].
  
\bibitem{Fukumi:2012rn}
  A.~Fukumi, S.~Kuma, Y.~Miyamoto, K.~Nakajima, I.~Nakano, H.~Nanjo, C.~Ohae and N.~Sasao {\it et al.},
  PTEP {\bf 2012} (2012) 04D002
  [arXiv:1211.4904 [hep-ph]].

\bibitem{Vissani:1999tu}
  M.~Doi, T.~Kotani, H.~Nishiura, K.~Okuda and E.~Takasugi,
  Phys.\ Lett.\ B {\bf 102} (1981) 323;
  F.~Vissani,
  JHEP {\bf 9906} (1999) 022
  [hep-ph/9906525];
  V.~Barger, S.~L.~Glashow, P.~Langacker and D.~Marfatia,
  Phys.\ Lett.\ B {\bf 540} (2002) 247
  [hep-ph/0205290];
  S.~Pascoli, S.~T.~Petcov and W.~Rodejohann,
  Phys.\ Lett.\ B {\bf 549} (2002) 177
  [hep-ph/0209059];
  A.~de Gouvea, B.~Kayser and R.~N.~Mohapatra,
  Phys.\ Rev.\ D {\bf 67} (2003) 053004
  [hep-ph/0211394].

\bibitem{Fogli:2004as}
  G.~L.~Fogli, E.~Lisi, A.~Marrone, A.~Melchiorri, A.~Palazzo, P.~Serra and J.~Silk,
  Phys.\ Rev.\ D {\bf 70} (2004) 113003
  [hep-ph/0408045];
  G.~L.~Fogli, E.~Lisi, A.~Marrone, A.~Melchiorri, A.~Palazzo, A.~M.~Rotunno, P.~Serra and J.~Silk {\it et al.},
  Phys.\ Rev.\ D {\bf 78} (2008) 033010
  [arXiv:0805.2517 [hep-ph]];
  M.~C.~Gonzalez-Garcia, M.~Maltoni and J.~Salvado,
  JHEP {\bf 1008} (2010) 117
  [arXiv:1006.3795 [hep-ph]].
  
\bibitem{Gando:2012zm}
  M.~Auger {\it et al.}  [EXO Collaboration],
  Phys.\ Rev.\ Lett.\  {\bf 109} (2012) 032505
  [arXiv:1205.5608 [hep-ex]].
  A.~Gando {\it et al.}  [KamLAND-Zen Collaboration],
  Phys.\  Rev.\  Lett.\  {\bf 110} (2013) 062502
  [arXiv:1211.3863 [hep-ex]].

\bibitem{Faessler:2013hz}
  A.~Faessler, G.~L.~Fogli, E.~Lisi, V.~Rodin, A.~M.~Rotunno and F.~Simkovic,
  arXiv:1301.1587 [hep-ph].
  
\bibitem{Arnaboldi:2002du}
  C.~Arnaboldi {\it et al.}  [CUORE Collaboration],
  Nucl.\ Instrum.\ Meth.\ A {\bf 518} (2004) 775
  [hep-ex/0212053];
  M.~Sisti [CUORE Collaboration],
  J.\ Phys.\ Conf.\ Ser.\  {\bf 203} (2010) 012069;
  F.~Bellini, C.~Bucci, S.~Capelli, O.~Cremonesi, L.~Gironi, M.~Martinez, M.~Pavan and C.~Tomei {\it et al.},
  Astropart.\ Phys.\  {\bf 33} (2010) 169
  [arXiv:0912.0452 [physics.ins-det]].
  
\bibitem{GomezCadenas:2011it}
  J.~J.~Gomez-Cadenas, J.~Martin-Albo, M.~Mezzetto, F.~Monrabal and M.~Sorel,
  Riv.\ Nuovo Cim.\  {\bf 35} (2012) 29
  [arXiv:1109.5515 [hep-ex]].
  
\bibitem{TheNEXT:2013lta}
  [ The NEXT Collaboration],
  arXiv:1307.3914 [physics.ins-det].
  
\bibitem{Lindner:2005kr}
  M.~Lindner, A.~Merle and W.~Rodejohann,
  Phys.\ Rev.\ D {\bf 73} (2006) 053005
  [hep-ph/0512143].
\end{thebibliography}
\end{document}